\documentclass[12pt,preprint]{aastex}
\usepackage{epsfig}
\usepackage{natbib}
\begin{document}
\title{Discovering Gravitational Lenses Through Measurements Of Their Time Delays}
\author{
Bart~Pindor,\altaffilmark{1,2}
}

\altaffiltext{1}{Department of Astronomy, University of Toronto, 60 St George Street, Toronto M5S 3H8, Canada}
\altaffiltext{2}{CFHT Legacy Survey Postdoctral Fellow}

\begin{abstract}
We consider the possibility that future wide-field time-domain optical imaging surveys may be able to discover gravitationally lensed quasar pairs through serendipitous measurements of their time delays. We discuss the merits such a discovery technique would have relative to conventional lens searches. Using simulated quasar lightcurves, we demonstrate that in a survey which observes objects several times each lunar cycle over the course of five years, it is possible to improve the efficiency of a gravitational lens search by 2-3 orders of magnitude through the use of time delay selection. In the most advantageous scenario considered, we are able to improve efficiency by a factor of 1000 with no loss of completeness. In the least advantageous scenario, we are able to improve efficiency by a factor of 110 while reducing completeness by a factor of 9. We show that window function effects associated with the length of the observing season are more important than the total number of datapoints in determining the effectiveness of this method. We also qualitatively discuss several complications which might be relevant to a real time delay search. 
 
\end{abstract}

\section{Introduction}

If the alignment of a distant quasar and the gravitational potential of a massive foreground galaxy causes multiple images of the quasar to be seen, then the quasar is said to be strongly lensed. Such systems can be utilized in a number of astrophysical investigations, including; modeling of the mass distribution of the lens galaxy \citep{1995ApJ...445..559K}, studying the ISM in lens galaxies \citep{1999ApJ...523..617F}, and constraining the source size through microlensing studies \citep{2000MNRAS.315...62W}. For a comprehensive review of the principles and applications of strong lensing, the reader is referred to \citet{2004kochanek_saasfee}. As stated by Kochanek, the most obvious first step in expanding the contributions of lensing-based investigations is to significantly increase upon the sample of $\sim$ 80 known lens systems. 

The ongoing Sloan Digital Sky Survey \citep[henceforth SDSS]{2000AJ....120.1579Y} will detect hundreds of thousands of optically-bright quasars. Given a canonical lensing rate of 1 in 1000, it should be possible to recover hundreds of lenses from the SDSS dataset. However, despite several successes (eg. Inada et al. 2003\nocite{Inada2}), the total number of new lenses discovered has been small.  

One of the main obstacles to recovering lenses from a survey such as the SDSS is the difficulty of obtaining follow-up observations to confirm, or refute, the lensing hypothesis for each lens candidate. Even in a lens search restricted to the SDSS spectroscopic sample, where the SDSS spectrum confirms the presence of at least one quasar image, each lens candidate typically requires high spatial-resolution spectroscopy, to confirm the presence of multiple quasar images with identical redshifts and similar spectral energy distributions (SEDs), and/or near-infrared imaging to confirm the presence of a lensing galaxy, before it can be confidently proclaimed to be a newly discovered lens system. Such observations can be difficult due to the small angular separations involved, and are generally expensive, in terms of telescope time, since they must be performed individually for each lens candidate.

Apart from components with very similar SEDs and the presence of a lensing galaxy, another identifying characteristic of a lensed quasar pair is the presence of a coherent time delay between the lightcurves of the two images. The time delay is the combined result of the geometric path length difference and the gravitational (often called the \citet{1964PhRvL..13..189}) delay. For a given lens system, a determination of the time delay can be considered the most unambiguous confirmation of the lensing hypothesis, as no other known mechanism can produce such a coherent delay. The possibility of measuring time delays, and the subsequent determination of the cosmological distance scale \citep{1964MNRAS.128..307R}, has been one of the long-standing motivations in the search for new lens systems. Unfortunately, measuring time delays requires frequent monitoring of the quasar images over an extended period of time and, for all but a handful of systems (eg. Kundic et al. 1997, Koopmans \& Fassnacht 1999, Burud et al. 2002, see Saha 2003 for a recent listing\nocite{1997ApJ...482...75K,1999ApJ...527..513K,2002A&A...391..481B,Saha03}), this effort has thusfar proven prohibitive. In general, the length of the delay depends on both the mass distribution of the lens and the position of the source, with typical galaxy-scale lenses producing delays in the range of 30-100 days.  

The increasing capabilities of CCD mosaic detectors (eg Gunn et al. 1998, Boulade et al. 2003 \nocite{1998AJ....116.3040G,2003SPIE.4841...72B}) have allowed for wide-field imaging surveys which combine extensive sky coverage, excellent image quality, and precise photometry. Recently, numerous proposals have been put forward to extend wide-field imaging into the time domain (eg \nocite{2002SPIE.4836...10T,2002SPIE.4836..154K,2000AIPC..540..263N}The Large-aperture Synoptic Survey Telescope (Tyson 2002), Pan-STARRS (Kaiser et al. 2002), Supernova / Acceleration Probe (Nugent 2000)) by repeatedly observing the same region of sky. The primary scientific motivations of such proposals include probing the dark energy through detection of supernovae, studying the distribution of dark matter through weak lensing, or detecting solar systems objects through their proper motions. Another merit of such time-domain observations is that any survey which repeatedly observes some area of the sky carries out a \textit{de facto} monitoring campaign of any lensed quasars in that survey area. Obviously, such a monitoring campaign could be used to measure the time delays of any known lenses in the survey area. In this work, we will consider a further possibility; that previously unknown systems could be identified as gravitational lenses through measurements of their time delays. 

\section{Detecting the Time Delay}

\subsection{Why Time Delays?}

To begin, let us briefly consider the merits of a selection technique based on detection of the time delay.

\textbf{Completeness:} Every lensed quasar pair must exhibit a time delay, but the completeness of the selection would be determined by whether or not that delay could be detected. Obviously, only delays shorter than the total survey duration could be detected. Time delays increase with increasing velocity dispersion of the lensing object and decrease with decreasing slope of the inner mass profile \citep{2003ApJ...583..584O}. Hence, a survey with a duration of several years will be sensitive to galaxy-scale lenses delays ($\leq$ 1yr), but unable to detect wider separation cluster-scale pairs. Further, since time delay measurements depend on comparing the lightcurves of lensed images, the source quasar has to vary measurably over the course of the survey. Most optically bright quasars do appear variable \citep{1996A&A...306..395C}, although the physical processes responsible for quasar variability are not well-understood. Fortunately, any survey which could measure time delays would also provide ample new empirical information on the variability of the entire quasar population. The completeness of the method will also be limited by the resolution at which close pairs can be accurately deblended, further discussed in \S \ref{simulations}. Finally, we discuss lens systems with more than two images in \S \ref{discussion}. 

\textbf{Efficiency:} Ideally, this method should have superb efficiency, as no other known astrophysical phenomenon produces a similar time delay in a pair of close point sources. In practice, time delay measurements are noisy and potentially misleading, even with excellent data. The efficiency of the selection will be determined by the extent to which the real time delays of gravitational lenses can be distinguished from coincidental correlations in the lightcurves of other point source pairs.

\textbf{Cost:} The measurement of a time delay is itself proof-positive of the lensing nature of a given lens candidate. Hence, it is possible that some lenses could be definitively identified from the survey data alone. This would be a significant improvement over existing optically-selected searches for which the rate of follow-up observations has thus far dominated the rate at which new lens systems are secured. 
 
\subsection{On Time Delays Statistics}

Before we begin to consider the conditions under which an object can be identified by its time delay, let us review how time delays are measured. A large and varied set of statistical methods has been proposed for measuring the time delay between a pair of lensed quasar images (eg. Vanderriest et al. 1989, Press, Rybicki \& Hewitt 1992 (PRH), Pelt et al. 1994, 1996, Pijpers 1997, Burud et al. 2001\nocite{1989A&A...215....1V,1994A&A...286..775P,1992ApJ...385..404P, 1997MNRAS.289..933P,2001A&A...380..805B}). Generically, these time delay statistics somehow combine the lightcurves of the two images to assign a figure-of-merit to each of a range of possible delays, with the measured delay being the one which is least excluded. However, as pointed out by Press et al., the lightcurves of \textit{any} two objects will produce \textit{some} favored (least excluded) time delay. Hence, the key to discovering gravitational lenses through their time delays is not simply to measure the value of the delay, but rather to demonstrate the significance of that measurement. It is difficult to decide \textit{a priori} which of the existing statistics would be best suited to this particular purpose. The choice is further complicated by the fact that the peculiarities of a given lightcurve and sampling rate can determine which statistic is best-suited to measuring the delay of that particular system \citep{2002A&A...381..428G}. 

We chose to use the dispersion statistic, $D^2_{4,2}$ invented by \citet{1996A&A...305...97P}, defined as

\begin{eqnarray}
D^{2}_{4,2} = \frac{\sum^{N-1}_{n=1} \sum^{N}_{m=n+1} S^{(2)}_{n,m} W_{n,m} G_{n,m} (C_n - C_m)^2}{\sum^{N-1}_{n=1} \sum^{N}_{m=n+1} S^{(2)}_{n,m} W_{n,m} G_{n,m}}
\end{eqnarray}

where

\begin{eqnarray}
S^{(2)}_{n,m} = \left\{\begin{array}{ll}
1 - \frac{|t_n - t_m|}{\delta}, & \mathrm{ if } \,|t_n - t_m| \leq \delta\\
0, & \mathrm{ if } \,|t_n - t_m| > \delta
\end{array} \right.
\end{eqnarray}

$C_i$ are the observed values for a combined lightcurve in which the points of one component have been time lagged by the putative delay value. $W_{n.m}$ is a statistical weight and $G_{n,m}$ is a term which ensures that only pairs of points originating from different components contribute to the sum (ie $G_{n,m} = 1$ if $C_n$ and $C_m$ are originally from different components and is zero otherwise). $S^{(2)}_{n,m}$ assigns a weight to each pair of points which linearly decreases for points further apart in time. We chose $\delta = 20$ days for the \textit{decorrelation length} parameter. Henceforth, the word 'dispersion' should be understood to refer specifically to the $D^2_{4,2}$ statistic. We chose this statistic because i) it is conceptually simple, ii) it can be rapidly calculated, and iii) it smooths the dispersion spectrum (figure-of-merit) in a manner which consolidates local minima produced by sampling peculiarities. Nonetheless, this choice is largely arbitrary and the question of which statistic is best suited for this purpose remains open.

\subsection{The Population of False Positives} 
\label{false_positives}

In order to determine the effectiveness of our proposed selection technique, we need to attempt to identify real lensed pairs from amongst a background population of plausible false positives. We could simply take this population to be every pair of close point sources in our theoretical survey, in which case pairs of stars would dominate the background population. In practice, any large wide-field survey would likely be conducted in multiple passbands, allowing quasar candidates to be efficiently identified on the basis of their broad-band colors \citep{2002AJ....123.2945R}. Further, before attempting to measure a time delay, we could reasonably require that both components of an object we are considering exhibit some appreciable level of variability. Any such variability threshold would further favor quasars over stars \citep{1989AJ.....98..108T,2002AcA....52..241E}. Together, these criteria would eliminate the vast majority of star plus star pairs and could largely discriminate against the quasar plus star superpositions which dominate the list of false positives in optical lens searches \citep{Pindor04}. Hence, we will take as our background population an ensemble of pairs of independent quasar lightcurves. This population can be thought of as corresponding to binary quasars, though there seem to be more lenses than binaries at small angular separations. More generally, this choice frames the problem as identifying pairs with time delays from amongst a sample of pairs not having time delays but with otherwise identical variability properties. An empirical approach to studying the population of possible false positives will be discussed in \S \ref{discussion}.     

\section{Monte Carlo Simulations} 
\label{simulations}
In this section we describe a set of Monte Carlo simulations which are designed to investigate the plausibility of the time delay selection method. We begin by describing the details of these simulations. 

\subsection{Simulation Method}

A conventional characterization of an objects variability is the structure function

\begin{eqnarray}
V(\Delta \tau) = \sqrt{\frac{\pi}{2}\langle|\Delta m (\Delta \tau)|\rangle^2}
\end{eqnarray}

which predicts the observed change in brightness, $\Delta m$, between observations separated by an interval of time, $\Delta \tau$. All of our simulated quasars were taken to be at a redshift of $z = 2$, and to have a rest-frame structure function of the form \citep{2004ApJ...601..692V}   

\begin{eqnarray}
V(\Delta \tau) = \left(\frac{\Delta \tau}{\Delta \tau_0}\right)^{\gamma}
\end{eqnarray}

where $\tau$ is measured in days, $\Delta \tau_0 = 5.36 \times 10^5, \gamma = 0.246$, as reported by \citet{2004ApJ...601..692V}. We constructed random pairs of lagged quasar lightcurves with the chosen structure function through the method prescribed by PRH. Photometric errors were taken to be 0.01 magnitudes throughout. One important caveat regarding photometry is that conventional object detection software does not deblend pairs of point sources whose separation is comparable to or less than the seeing FWHM. \citet{2003AJ....125.2325P} demonstrated that, through direct modeling using an empirical PSF, it is possible to separate pairs at considerably smaller separations, but, for SDSS imaging of the median SDSS quasar, the errors associated with this modeling would dominate the photometric uncertainty in the fluxes of the components. Hence, we have assumed that our theoretical survey will have sufficient depth and image quality that the majority of lens candidate pairs can be deblended with an accuracy which does not dominate the photometric error. The validity of this assumption, or equivalently the geometrical selection function, for any real dataset could be ascertained by an appropriate set of image simulations. We assumed that the total length of the survey is five years, comparable to existing surveys such as the SDSS and the Canada-France-Hawaii Telescope Legacy Survey (CFHTLS)\footnote{http://www.cfht.hawaii.edu/Science/CFHTLS/}. The final detail of our simulated lightcurves is the sampling rate. We devote the remainder of this section to illustrating the effects of different sampling rates on the selectibility of lensed pairs.

\subsection{Constant Sampling}

\begin{figure}
\plotone{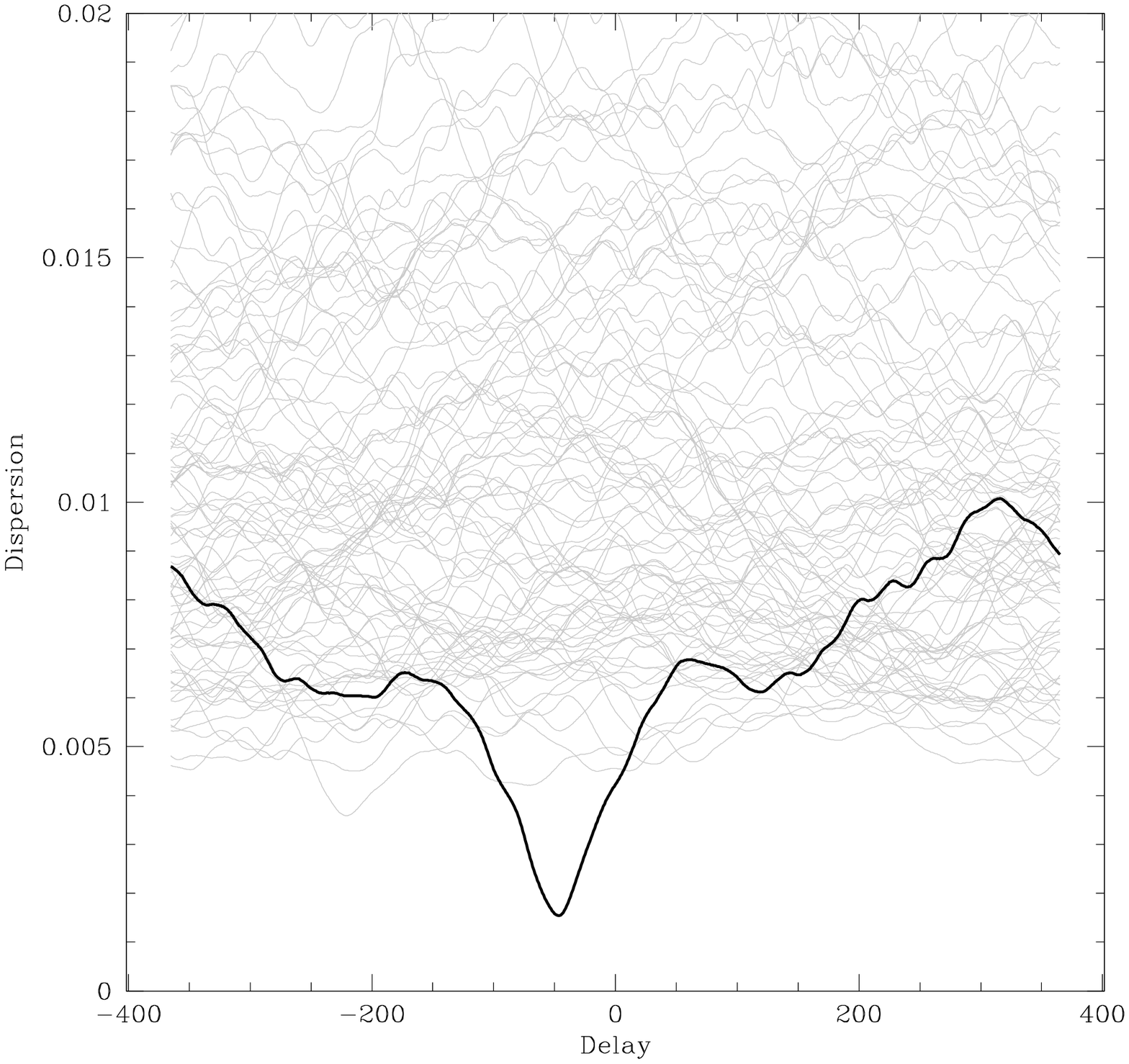}
\caption{The dispersion spectrum for two images of a simulated lensed quasar (in black) and 100 simulated independent quasar pairs (in grey). The simulated lens has a constructed delay of -45 days.}
\label{ideal_dispersion}
\end{figure}

The first possibility we considered was the case of constant sampling, For this case, we assume that objects are observed every five days, throughout the length of the survey. Clearly, such a sampling rate is unrealistic for an optical survey, but the purpose of this example is to illustrate how a real lens would be differentiated from a population of false positives in the ideal case, and it will be instructive to compare this case to the more realistic sampling rates we shall consider later. Figure \ref{ideal_dispersion} shows the dispersion spectrum for a lensed quasar pair as well as for 100 independent comparison pairs. Here, and in all subsequent figures, we restricted the range of considered delay values to be $|\tau| < 365$ days.

Of course, in a real lens search, all of the curves in this figure would be the same color, but, even so, it would not be difficult to distinguish the dispersion spectrum corresponding to the lensed pair. The real lens has an obvious and well-defined dispersion minimum with a value significantly lower than any of the comparison minima. 

\subsection{Sampling Rates and the Window Function}

The sampling rate of a real optical survey is restricted by limited observing seasons, the lunar cycle, bad weather, and mechanical failures. Uneven sampling can greatly affect the effectiveness with which certain delays can be excluded through statistical tests. This varying sensitivity caused by the sampling rate is often referred to as the \textit{window function}. The window function has two important effects: first, it can cause the time delays of real lens pairs to be inaccurately or incorrectly measured, and second, it can cause pairs of independent lightcurves to systematically pick out certain time delay values. One of the merits of the method we are considering is that, as long as our lens candidates are observed with wide-field detectors, there will always be a large number of other objects in the field observed with the same sampling rate. This ensemble of comparison object will reveal whether or not the sampling rate produces time delay signatures which cannot be distinguished from the time delay of a real lensed pair. 

We investigated the effects of the window function by simulating a number of plausible sampling rates which might be appropriate for a future facility such as the LSST; an instrument capable of observing a large fraction of the available sky every few days. We considered a high sampling rate, with objects being observed 5 to 7 times within $\pm$10 days of each new moon, and a low sampling rate, 2 to 4 times within $\pm$7 days. In both cases, observations are made not less than two days apart. For both the low and high sampling rates, we considered observing seasons of 6 and 8 months.  

\begin{figure}
\plotone{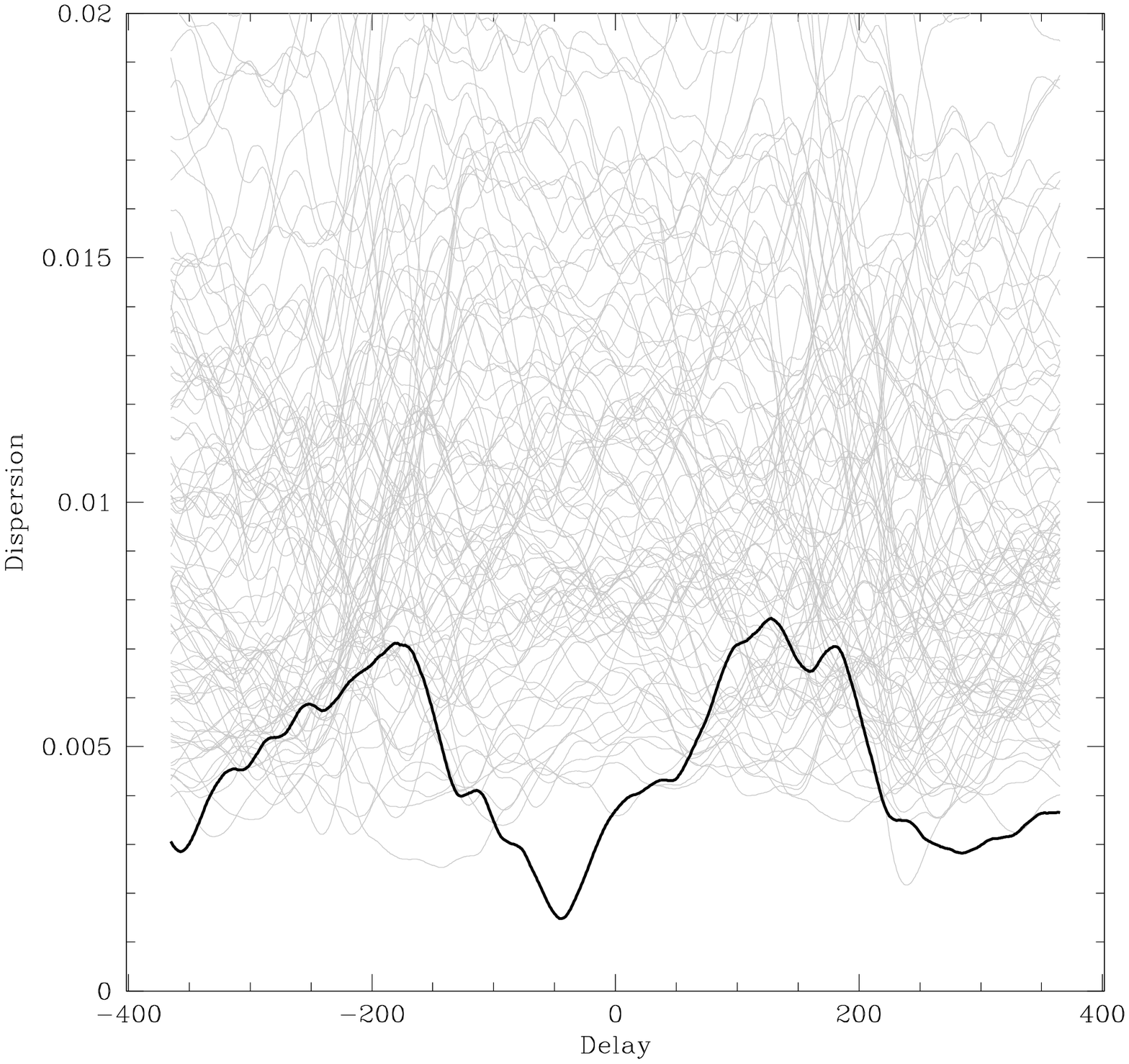}
\caption{The dispersion spectrum of a simulated lensed quasar pair at the high sampling rate (see text) and 8 month observing seasons (in black) and 100 comparison pairs (in grey). The simulated lens has a constructed delay of -45 days.}
\label{single_8months}
\end{figure}

\begin{figure}
\plotone{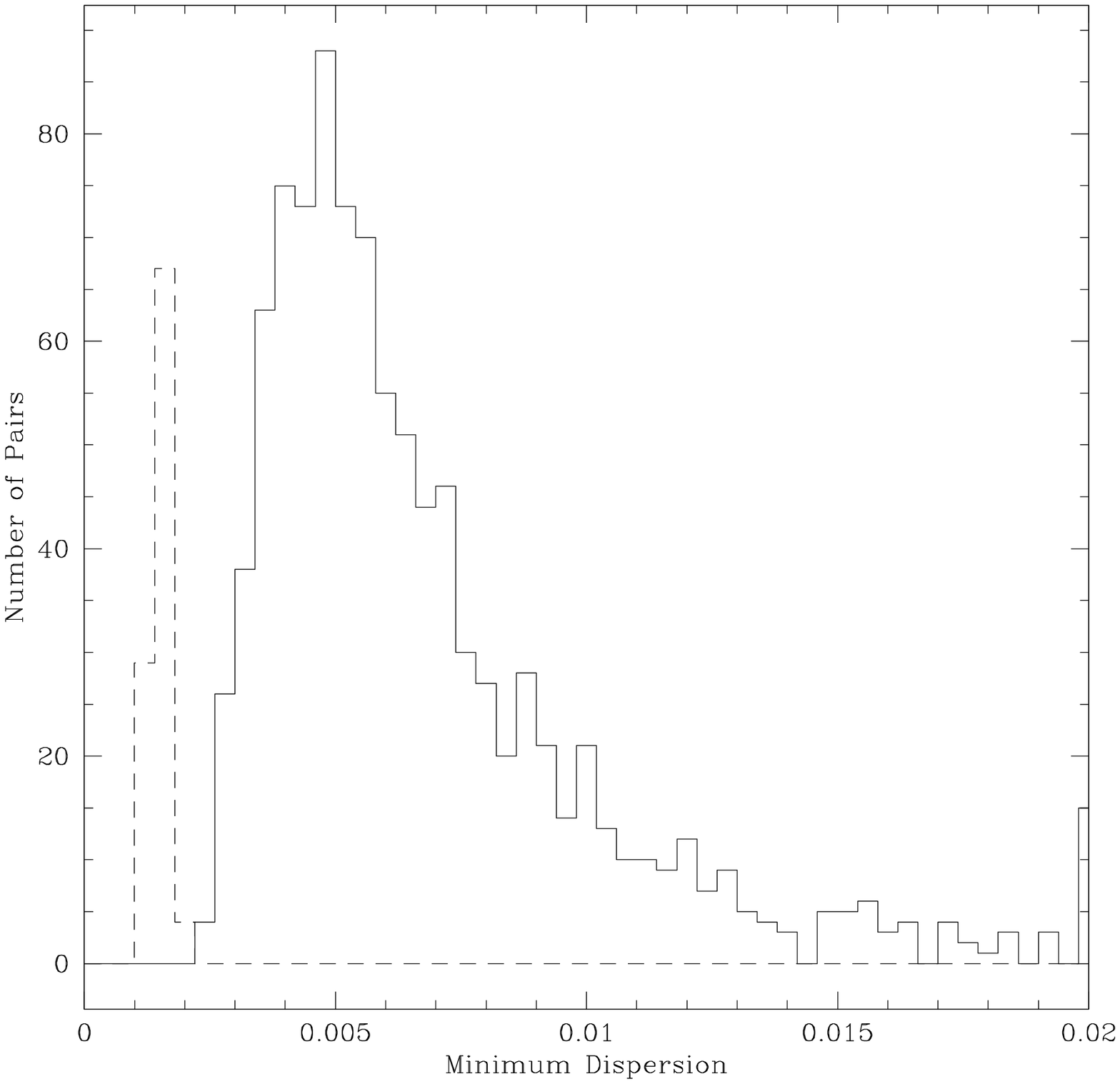}
\caption{Histograms indicating the value of the minimum dispersion for 100 lensed pairs (dashed line) and 1000 comparison pairs (solid line).}
\label{8months_dmins_hist}
\end{figure}

Figure \ref{single_8months} shows the dispersion spectrum of a real lensed pair, together with 100 comparison pairs, observed at the high sampling rate with 8 month observing seasons. As in the constant sampling scenario, the real lens can be identified as having the lowest dispersion minimum. We constructed a larger simulated ensemble to demonstrate that this is generally true at this sampling rate. We simulated 100 lensed pairs with independent realizations of the sampling rate and having time delays randomly chosen to be between -200 and 200 days. The distribution of real time delays would not be uniform (eg.consult figure 5 of \citet{2002ApJ...568..488O}), but a uniform distribution best illustrates the effectiveness of this selection method itself.  For each lensed pair, we created 10 comparison independent quasar plus quasar pairs. Figure \ref{8months_dmins_hist} shows the distribution of dispersion minima for both the lensed and comparison pairs. The two populations separate cleanly, and if we make a selection by requiring that the minimum dispersion be less than 0.0025, then we recover all 100 lenses and only 3 (out of 1000) comparison pairs. If we generate a similar ensemble at the low sampling rate with 8 month observing seasons, and apply the same selection, then we recover 99 lenses and 19 comparison objects. Table \ref{efficiency_table} shows the relative efficiency of various selection criteria.

\begin{figure}
\plotone{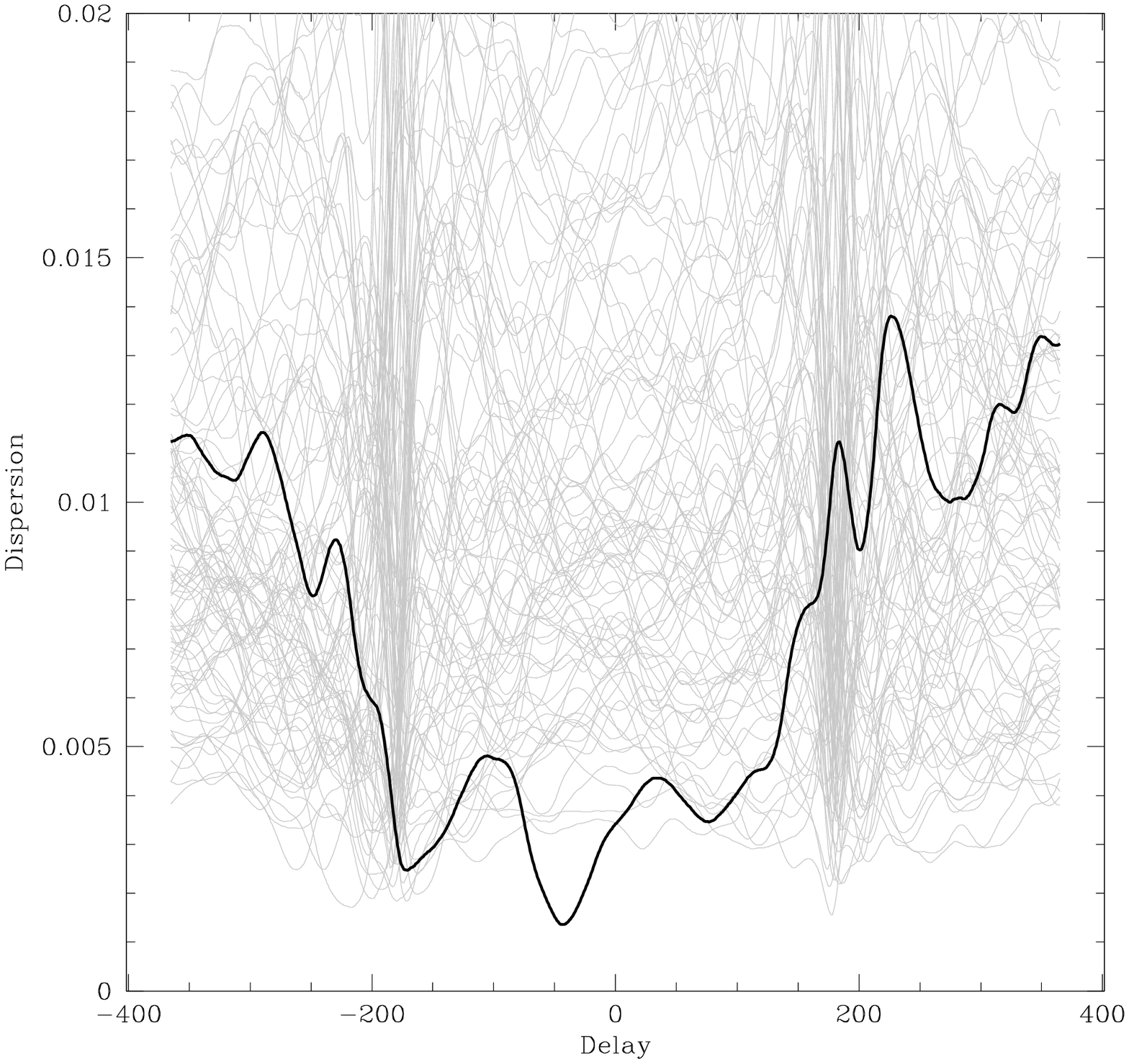}
\caption{The dispersion spectrum of a simulated lensed quasar pair at the high sampling rate (see text) and 6 month observing seasons (in black) and 100 comparison pairs (in grey). The simulated lens has a constructed delay of -45 days.}
\label{single_6months}
\end{figure}

\begin{figure}
\plotone{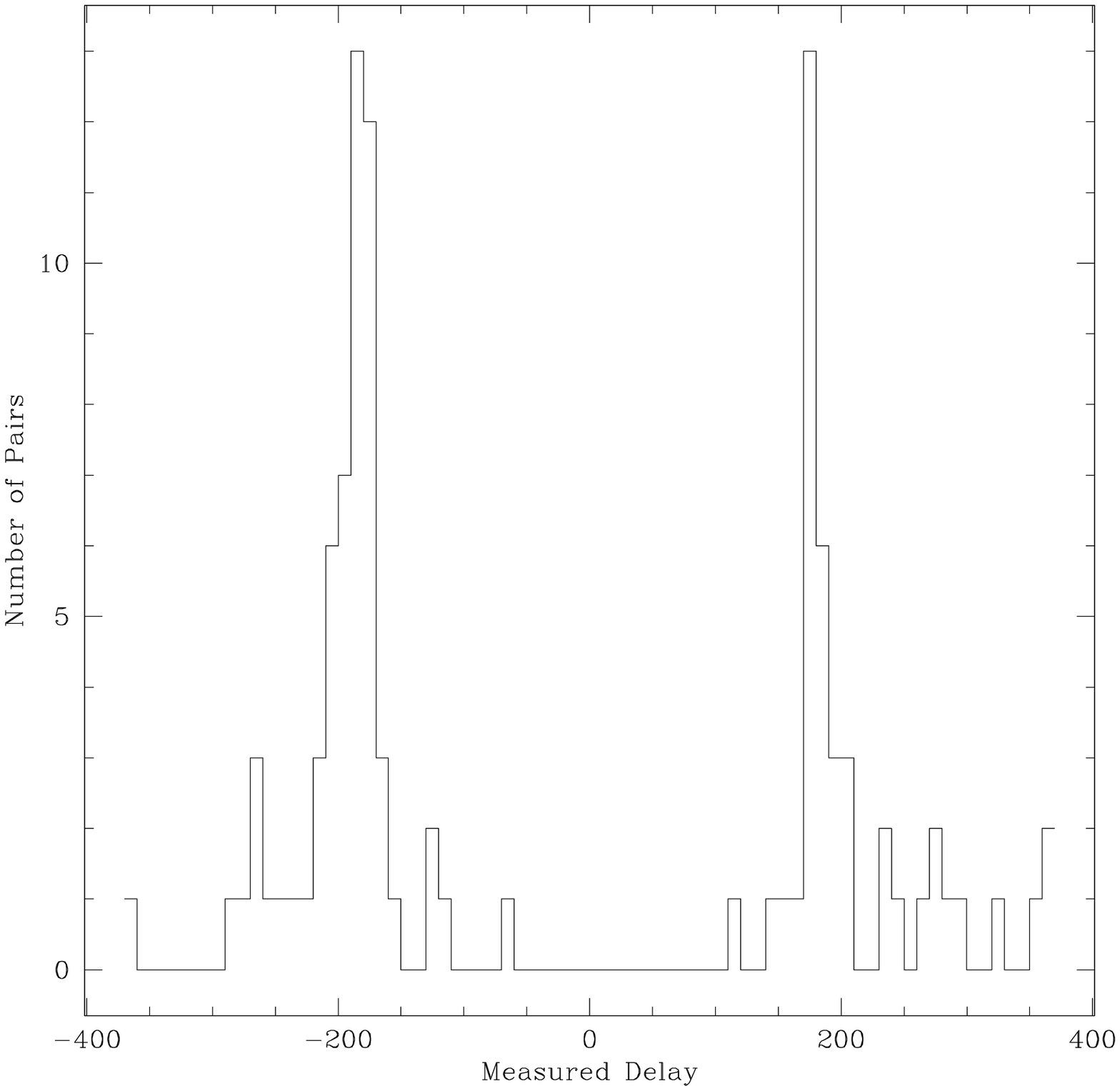}
\caption{Histograms indicating the value of the minimum dispersion for 100 lensed pairs (dashed line) and 1000 comparison pairs (solid line).}
\label{6months_dmins_hist}
\end{figure}

When we consider 6 month observing seasons, the effects of the window function become much more prominent. Figure \ref{single_6months} is analogous to figure \ref{single_8months}, but with a 6 month observing season. We can once again identify the lensed pair as having the lowest dispersion minimum, but in this case of greater interest is the behavior of the comparison spectra in the regions of $\sim \pm 180$ days. At these delay values, the number of observations contributing to the dispersion (ie for which $S^{(2)}_{n,m} \neq 0$) is small, making the statistic very noisy. Figure \ref{6months_dmins_hist} shows the distribution of measured delays (location of minima) for the 100 comparison objects shown in figure \ref{single_6months}. Clearly, there is a systematic tendency for objects to exhibit dispersion minima at delays associated with the window function. This effect can both aid and hinder our time delay selection. On the one hand, if we consider our previous minimum dispersion threshold, we find that, from an ensemble of again 100 lenses and 1000 comparison objects, we recover 99 lenses and 284 comparison objects. For the low sampling rate, we recover 100 lenses and 577 comparison objects. Clearly, window function effects can produce dispersion minima which have values comparable to those produced by real time delays, degrading the selection efficiency. On the other hand, these false minima systematically occur at particular delay values. We can increase the efficiency of our selection by excluding objects with measured delays heavily contaminated by the window function. For instance, if we require that $|\tau| < 150$, then, in the high sampling case, we recover 63 lenses and 7 comparison object and, in the low sampling case, 37 lenses and 12 comparison objects. Of course, restricting the range of allowed delays excludes some lensed pairs by construction, but note that the reduced completeness is also partly due to the exclusion of lensed pairs whose time delays are incorrectly measured because of the window function.     

Given that we have had to speculate regarding the capabilities of a future instrument, the results of this section should not be considered specific predictions of the effectiveness of an actual time delay selection. Instead, we take these results to generally confirm the plausibility of this selection method. The length of the observing season appears to be the most significant factor in determining the effectiveness of this method.  

\subsection{Identifying Real Delays}

Existing time delay statistic were designed with the goal of accurately measuring the time delay of a pair of objects which are known, or at least strongly suspected, to be lensed images of the same source. In the case of time delay selection, our primary goal is to distinguish real delays from coincidental alignments. Certainly, the results of the previous section show that existing statistics can be used quite effectively to this end. Nonetheless, it is interesting to ask if criteria specifically designed to identify real delays can improve the effectiveness of our selection.

 Suppose that, for each object, we were to divide our set of observational epochs into two equal and distinct subsets. For a real lensed pair, because each of these subsets ``encode'' the same time delay as the full dataset, we would expect that each would produce the same, if somewhat noisier, measured delay. However, for one of our comparison objects, whose time delay minima are solely the result of coincidental noise, we would expect the two independent subsets, absent window function effects, to produce two completely unrelated time delay minima. In this sense, the lightcurve of a real lensed pair produces a robust delay measurement. We now demonstrate a simple procedure to quantify this robustness. 

\begin{figure}
\plotone{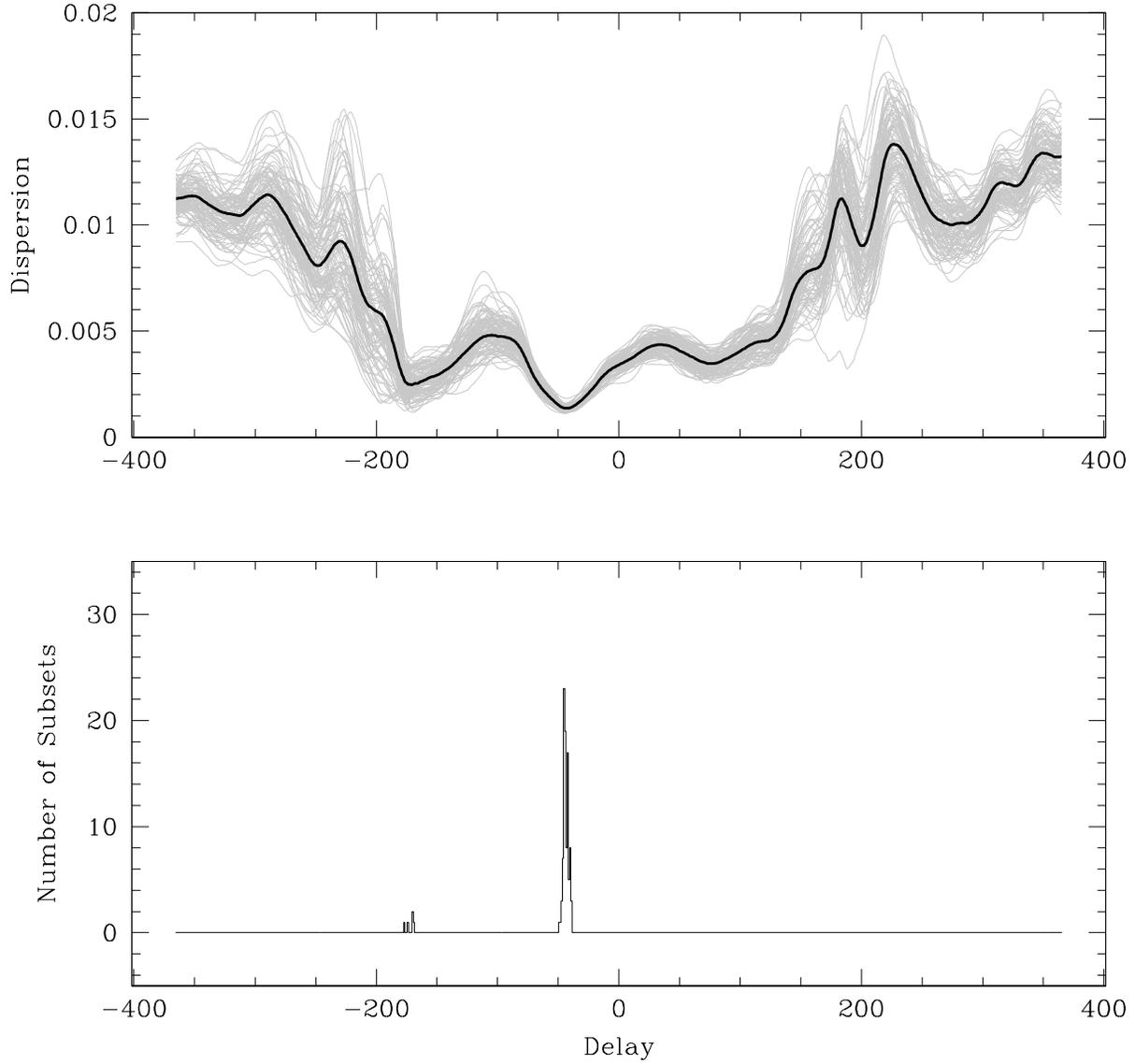}
\caption{Upper Panel: The dispersion spectrum (in black) of a simulated lensed quasar pair, together with the dispersion spectra (in grey) of 100 jackknife subsets. Lower Panel: Histogram of the dispersion minima for the 100 subsets shown in the upper panel.}
\label{jackknife_lens}
\end{figure}

\begin{figure}
\plotone{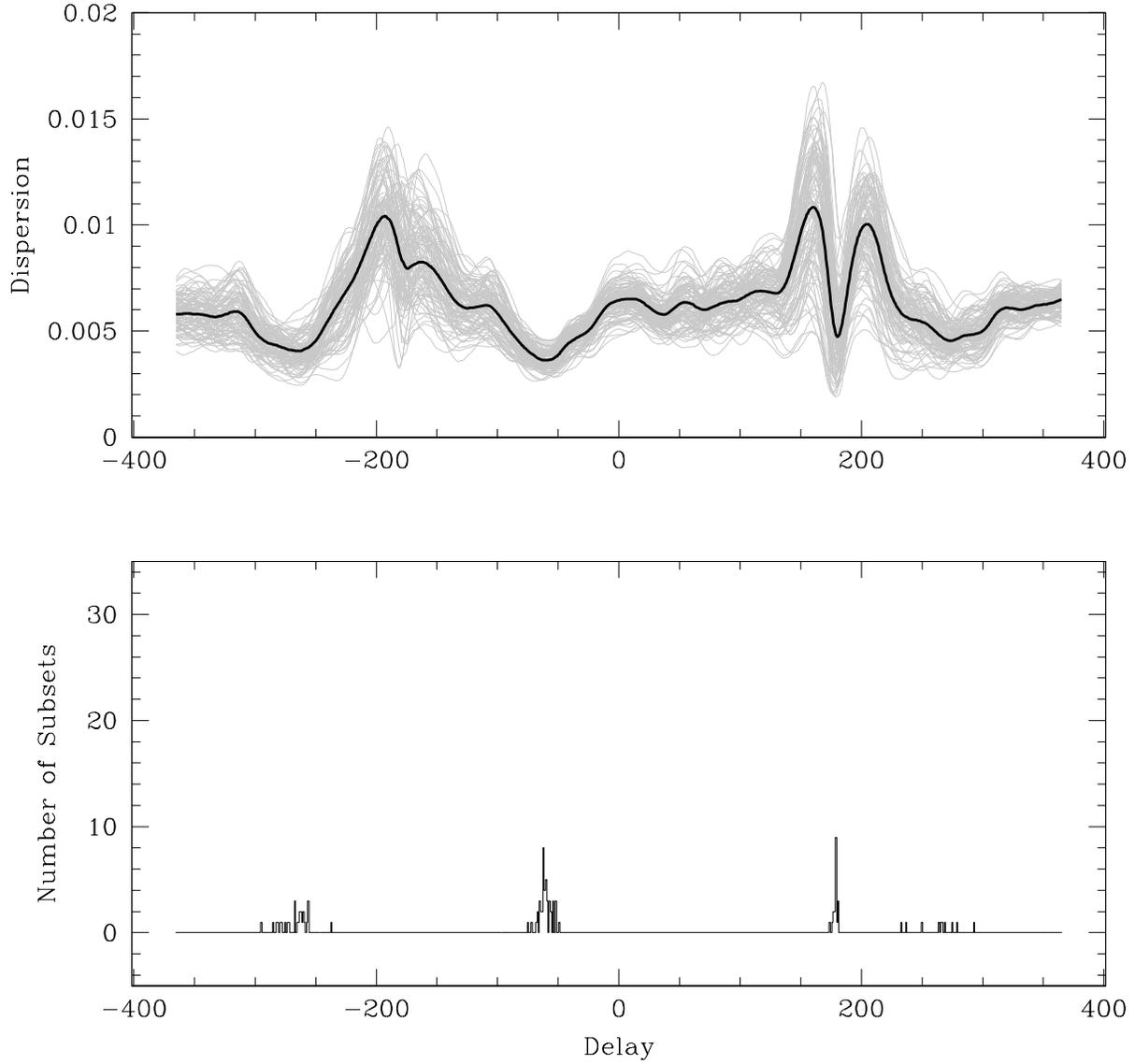}
\caption{Upper Panel: The dispersion spectrum (in black) of a comparison object, together with the dispersion spectra (in grey) of 100 jackknife subsets. Lower Panel: Histogram of the dispersion minima for the 100 subsets shown in the upper panel.}
\label{jackknife_comparison}
\end{figure}

Only two of the objects shown in figure \ref{single_6months} have measured delays of less than 100 days; one is the lensed pair and one is a comparison object. For each of these two objects, we created 100 subsets by randomly omitting 37 \% ($1/e$) of the data points in the original lightcurve. Figure \ref{jackknife_lens} shows the dispersion spectrum of the lensed pair from the full dataset together with the dispersion spectra of 100 subsets. Also shown is the distribution of measured time delays for the 100 subsets. As predicted, the majority (94) of the subsets produce measured time delays which are very close (within 5 days) to the value measured for the full dataset. In constrast, figure \ref{jackknife_comparison} shows the dispersion spectrum of the comparison object, together with the dispersion spectra of 100 similiar subsets. In this case, only 31 of the subsets have measured delays within 5 days of the original measured value. We used the semi-interquartile range (SIQR) of the measured subset delays to quantify this distinction; if the SIQR is small, then the majority of subsets have indicated the same delay. In such a case, this procedure is essentially equivalent to error estimation by jackknife resampling \citep{1982jbor.book.....E}. However, if the SIQR is large, then it does not correspond to a well-defined error, but instead implies that the data do not convincingly indicate a single delay value.  

As shown in Table \ref{efficiency_table}, introducing the selection criterion SIQR $<$ 5 does improve the selection efficiency for each sampling rate. The efficiency gains are, however, fairly modest and, in the worst sampling scenario, involve a substantial loss in completeness. Of course, we should have preferred that the SIQR would be a more dramatic improvement in our selection methodology, however our primary goal in the section has been to simply bring attention to the possibility of creating a statistical measure specifically designed to distinguish between real and coincidental measured delays.  

\section{Discussion}
\label{discussion}

In this section we consider qualitatively several subjects which are beyond the scope of our lightcurve simulations, but which are nonetheless relevant to the time delay selection method.

\textbf{Additional Variability Selection:} A major contaminant in existing optical lens searches are quasar plus star superpositions. As mentioned in \S \ref{false_positives}, such candidates could often be eliminated through variability, or even proper motion \citep{2001A&A...373...38B}, cuts. The effectiveness of such selection techniques will be well-known through application to individual quasars in the survey.

\textbf{Realistic Quasars and Backgrounds:} The selection efficiencies we have estimated might be grossly inaccurate if the form we have assumed for quasar lightcurves badly misrepresents their relevant variability properties, or if there is a population of background objects which are much more likely to be mistaken for lensed pairs than the independent quasar pairs we have used. Fortunately, it will be soon be possible to empirically test these possibilities. The deep component of the CFHTLS will image 4 square degrees of the sky more than 100 times over 5 years. Although this area is not sufficient to expect many lensed pairs, the data will provide ample information on relevant quasar variability, and will allow for the identification of any unexpected yet common class of false positives. Further, any lens candidate selected through this method will be observed in a field with thousands of other point sources. By calculating delay statistics for random pairings of these field objects, it will be possible to ascertain the significance of any candidate. 

\textbf{Microlensing:} One complication we have ignored in creating our simulated lightcurves is the possibility of microlensing of the lensed quasar images. As pointed out by Pelt et al., microlensing variations violate several of the statistical assumptions implicit in the lightcurve generating procedure of PRH which we have employed. Further, it is only real lenses, and not background objects, which can be expected to exhibit an appreciable rate of microlensing. The presence of microlensing variations would undoubtedly make it more difficult to correctly identify the time delays of lensed pairs (although such variations might make it easier to identify lenses through other variability criteria), but we have not attempted to estimate how serious this effect might be.  

\textbf{Faint Lenses:} The imaging surveys we have considered will obtain accurate photometry for large numbers of lensed quasars well below the break of the optical quasar luminosity function. The time delay selection method could conceivably recover such faint lenses. In fact, the anti-correlation between quasar luminosity and variability found by Vanden Berk et al. suggests that fainter quasars might be favored. One possible complication is flux from the lens galaxy; as the quasar images become relatively fainter, correcting the quasar lightcurves for lens galaxy flux (for instance through image subtraction) will be required. We might note that detection of the lens galaxy in addition to the variable quasar images could itself be an indication of lensing. Certainly, for ever-fainter quasars, lensing confirmation through time delays becomes ever more attractive when compared to the prospect of obtaining spatially-resolved spectra.   

\textbf{Four Image Lenses:} Throughout this work, we have only considered to two image gravitational lens systems whereas, in reality, some $\sim 25-50 \%$ of lensed quasars exhibit four images \citep{2001ApJ...553..709R,2003MNRAS.341...13B,Cohn03}. Rather than being an oversight, this is simply because the discovery method described in this paper is essentially superfluous for such lenses; a single image which resolves the different components of a four image lens is sufficient evidence for a confident identification of that object as a likely lens systems. Two image lenses must be distinguished from the large population of chance point source alignments, but the corresponding population of false positives for four image lenses, a chance superposition of four point sources of comparable brightness within 1-2$^{\prime\prime}$ of one another, is virtually non-existent. 

\textbf{The Lengths of Delays and Observing Seasons:} As we have demonstrated, it is easier to select lenses whose delays are considerably shorter than the length of the observing season. Fortunately, as stated, galaxy-scale lenses are both expected and seen to have delays typically in the range of 30-100 days. On the other hand, it is reasonable to expect that the strategy of future surveys will partially be dictated by the desire to improve image quality by reducing airmass, thus reducing the effective length of the observing season for many objects. Finally, the distribution of observing seasons for objects across the sky will be determined by the latitude of any ground-based site.     

\section{Conclusions}

We have proposed that future imaging surveys which repeatedly observe the same region of sky may be able to discover previously unknown gravitationally lensed quasar pairs through measurements of their time delays. We have applied a series of simple selection criteria to predict that use of time delay information should improve the efficiency of a gravitational lens search by 2-3 orders of magnitude, while generally maintaining high completeness. The selection procedure we have proposed should be considered illustrative, with the best possible method for identifying time delays remaining a matter open to investigation. Our simulations indicate that window function effects associated with the length of observing seasons are the most serious limitation to time delay detections. In the future, improved understanding of quasar variability, empirical determination of the population of possible false positives, and more specific information regarding the characteristic of future time domain surveys will make it possible to further confirm whether or not detection of time delays will be an important element of future gravitational lens searches.  

My thanks to Ed Turner for his numerous helpful suggestions. This work has been made possible through the support of the Natural Sciences and Engineering Research Council of Canada.

\clearpage

\begin{deluxetable}{lcc}
\tablecolumns{3}
\tablewidth{0pc}
\tablecaption{Summary of Selection Results}
\tablehead{
\colhead{Selection} & \colhead{Number of} & \colhead{Number of} \\
\colhead{Criterion} & \colhead{Lensed Pairs} & \colhead{Comparison Objects} \\
\colhead{} & \colhead{Selected} & \colhead{Selected} 
}
\startdata
\textbf{No Selection} & 100 & 1000 \\
\cutinhead{8 Month Observing Season - High Sampling Rate}
\textbf{$D^{2}_{4,2} <$ 0.0025} & 100 & 3 \\
\textbf{$D^{2}_{4,2} <$ 0.0025 and SIQR $<$ 5} & 100 & 1 \\
\cutinhead{8 Month Observing Season - Low Sampling Rate}
\textbf{$D^{2}_{4,2} <$ 0.0025} & 99 & 19 \\
\textbf{$D^{2}_{4,2} <$ 0.0025 and SIQR $<$ 5}  & 94 & 9 \\
\cutinhead{6 Month Observing Season - High Sampling Rate}
\textbf{$D^{2}_{4,2} <$ 0.0025} & 99 & 284 \\
\textbf{$D^{2}_{4,2} <$ 0.0025 and Delay $<$ 150 days} & 63 & 7 \\
\textbf{$D^{2}_{4,2} <$ 0.0025 and Delay $<$ 150 days and SIQR $<$ 5}  & 50 & 2 \\
\cutinhead{6 Month Observing Season - Low Sampling Rate}
\textbf{$D^{2}_{4,2} <$ 0.0025} & 100 & 577 \\
\textbf{$D^{2}_{4,2} <$ 0.0025 and Delay $<$ 150 days} & 37 & 12 \\
\textbf{$D^{2}_{4,2} <$ 0.0025 and Delay $<$ 150 days and SIQR $<$ 5}  & 11 & 1 \\
\enddata
\label{efficiency_table}
\end{deluxetable}

\end{document}